\documentclass[a4paper,11pt]{article}

\usepackage{jheppub} 

\usepackage{enumitem}
\usepackage{mathrsfs}  
\usepackage[normalem]{ulem}

\title{Topology and $\theta$ dependence in finite temperature $G_2$ lattice
       gauge theory}
\author{Claudio Bonati} 
\affiliation{INFN - Sezione di Pisa,\\ Largo Pontecorvo 3, I-56127 Pisa, Italy}
\emailAdd{bonati@pi.infn.it}

\abstract{In this work we study the topological properties of the $G_2$ lattice
gauge theory by means of Monte Carlo simulations. We focus on the behaviour of
topological quantities across the deconfinement transition and investigate
observables related to the $\theta$ dependence of the free energy. As in
$SU(N)$ gauge theories, an abrupt change happens at deconfinement and an
instanton gas behaviour rapidly sets in for $T>T_c$.}


\arxivnumber{1501.01172}

\begin{document} 
\maketitle
\flushbottom

\section{Introduction}
\label{sec:intro}

A large part of the standard particle phenomenology depends on phenomena like
color confinement and chiral symmetry breaking, however a complete
understanding of these nonperturbative phenomena is still lacking. In order to
better figure out  these phenomena, a common strategy has been to extend the
theory under study beyond its natural setting, in order to find regularities or
more general patterns underlying the usual structure. 

As an example, real life Quantum Chromodynamics (QCD) is a theory with gauge group
$SU(3)$ and fermions with particular mass values, however, in order to grasp
some intuition on its strongly coupled dynamics, it is convenient to look at
theories with fermions of arbitrary masses, with particular attention to the
extremal cases of massless fermions and pure $SU(3)$ theory. In the same way,
it is convenient to think of the $SU(3)$ theory just as a particular realization
of the general theory with gauge group $SU(N)$, which is, e.g., the natural
setting for techniques like the large $N$ expansion.  
Although these approaches were not capable of providing quantitatively reliable
predictions, their qualitative indications are of the utmost importance to get
some understanding of the QCD nonperturbative physics. 

An even more drastic extension is obtained by considering gauge theories with
an arbitrary gauge group. In this paper we will study some properties of the
theory with the exceptional group $G_2$ as gauge group. In order to motivate
this apparently bizarre choice we need some background. 

In pure gauge theory, at temperature $T$, the free energy
$F_{Q\bar{Q}}(\vec{r}\,)$ of a couple of static color-anticolor charges at distance
$\vec{r}$ is given by the expression \cite{McLerran:1981pb} 
\begin{equation}\label{eq:FQbarQ}
\exp\left(- F_{Q\bar{Q}}(\vec{r}\,)/ T\right)=\langle P(0)P(\vec{r}\,)^{*}\rangle\ ,
\end{equation}
where $P(\vec{r}\,)$ is the Polyakov loop. In the Lattice Gauge Theory 
(LGT, \cite{Wilson:1974sk}) setting this can be written as
\begin{equation}\label{eq:poly}
P(\vec{r}\,)=\mathrm{Tr}\prod_{t=0}^{N_t-1}U_0(\vec{r},t)\ ,
\end{equation}
the $U_{\mu}(\vec{r},t)$s being the elementary parallel transports along the
links of the lattice and $N_t$ the number of lattice elements in the
compactified temporal direction. From Eq.~\eqref{eq:FQbarQ} it follows that a
couple of static color charges can be separated at arbitrary distances only if
the thermal average of the Polyakov loop is different from zero: $\langle
P\rangle \neq 0$.  Let us now consider for every $\vec{r}$ the following
transformation
\begin{equation}\label{eq:center}
U_0(\vec{r}, \bar{t})\to U_0'(\vec{r},\bar{t}\,)=Z U_0(\vec{r},\bar{t}\,)\ ,
\end{equation}
where $\bar{t}$ is a fixed time slice of the lattice, and $Z$ is an element of
the center of the gauge group (independent of $\vec{r}$\,). It is simple to
show that the Wilson lattice action (see Sec.~\ref{sec:G_2}) is invariant under the
transformation Eq.~\eqref{eq:center}, while clearly the average of the Polyakov
loop transforms as $\langle P\rangle \to Z\langle P\rangle$. As a consequence
deconfinement can be associated to the spontaneous breaking of the symmetry
Eq.~\eqref{eq:center}, i.e. of the center symmetry (see e.g.
\cite{Greensite:2003bk} for much more details).

Although this picture of confinement appears quite satisfying and leads to
nontrivial predictions (like e.g. the Svetitsky-Yaffe conjecture
\cite{Svetitsky:1982gs} on the universality class of the deconfinement
transition), it clearly does not cover two class of theories:
\begin{enumerate}
\item theories whose action is not invariant under Eq.~\eqref{eq:center}, 
      like theories with fermions in the fundamental color representations;
\item theories with a trivial gauge group center.
\end{enumerate} 

Theories with fermions are notoriously computationally demanding, so they are
not the first choice for a study that, depending on the observable to be
monitored, could require high statistics. Concerning the theories of the second
class, the simplest group with trivial center that comes to mind is $SO(3)$,
however in this case (and more generally for all the $SO(N)$ groups) large
lattice artefacts make a systematic lattice investigation problematic, see e.g.
\cite{deForcrand:2002vs}. A particularly interesting alternative proved to be
the gauge group $G_2$.  Beyond having trivial center, $G_2$ presents two
peculiar features: 
\begin{enumerate}[label=\alph*)]
\item it is simply connected;
\item a charge in the fundamental representation ($\mathbf{7}$) can be screened 
by charges in the adjoint representation ($\mathbf{14}$).
\end{enumerate}

The first of these properties is interesting when studying confinement, since
it means that a $G_2$ gauge theory does not support topologically stable vortex
configurations, thus at least one of the possible models of color confinement
(i.e. the vortex picture, see e.g. \cite{Greensite:2003bk}) requires
non-trivial modifications to be applied in the $G_2$ setting.  For this reason,
together with the fact that $G_2$ has the low rank value $2$, the $G_2$ gauge
theory was often used as a testbed for possible confinement mechanisms (see
e.g. \cite{Poppitz:2012nz, Anber:2014lba, Diakonov:2010qg, Bonati:2010tz,
DiGiacomo:2008gf}).  The second of the aforementioned properties\footnote{which
follows from the fact that the Clebsch-Gordan series of the product
$\mathbf{7}\otimes \mathbf{14}\otimes\mathbf{14}\otimes \mathbf{14}$ contains a
singlet, see e.g.  \cite{Holland:2003jy}.} makes the $G_2$ pure gauge theory
quite similar to a theory coupled to matter, in which color is confined but the
area law for the Wilson loop is not valid (or, equivalently, the asymptotic
string tension vanishes) because of the string breaking phenomenon.

Bearing all this in mind, it is not surprising that the $G_2$ lattice gauge
theory was actively investigated in the past, both at zero and finite
temperature \cite{Holland:2003jy, Pepe:2005sz, Greensite:2006sm, Pepe:2006er,
Cossu:2006bi, Cossu:2007dk, Maas:2007af, Liptak:2008gx, Danzer:2008bk,
Wellegehausen:2010ai, Ilgenfritz:2012wg, Bruno:2014rxa}. What is probably
surprising is that the results of these analysis gave a picture much similar to
that of standard $SU(N)$ theory: the spectrum of the $G_2$ theory at zero
temperature is composed only of color neutral objects \cite{Holland:2003jy,
Pepe:2005sz, Cossu:2006bi}, the string tension at intermediate distances (i.e.
before string breaking \cite{Wellegehausen:2010ai}) satisfies Casimir scaling
\cite{Greensite:2006sm, Liptak:2008gx, Wellegehausen:2010ai}, a first order
deconfinement transition is present \cite{Pepe:2006er, Cossu:2007dk,
Bruno:2014rxa}, (quenched) chiral symmetry is broken in the low temperature
phase and restored above the critical temperature \cite{Danzer:2008bk}, the
topological susceptibility is suppressed above deconfinement
\cite{Ilgenfritz:2012wg}, propagators \cite{Maas:2007af} and thermodynamical
observables (like e.g. pressure and trace anomaly) \cite{Bruno:2014rxa} do not
show any qualitative difference with respect to the $SU(N)$ case.

In this paper we will study the topological properties of the $G_2$ LGT near
the deconfinement transition, with particular attention to observables related
to the functional dependence of the free energy on the $\theta$ angle. As will
be recalled in more detail in the next section, in $SU(N)$ one expects an
abrupt change of the functional form of the free energy at deconfinement,
switching from the large $N$ behaviour in the low-$T$ phase to an instanton gas
behaviour in the high-$T$ phase. It is \emph{a priori} clear that such an
argument, based on the large $N$ analysis of $SU(N)$ gauge theories cannot be
directly applied to the case of the $G_2$ LGT, nevertheless our results
indicate that also in this case the $G_2$ theory resembles very much the
$SU(N)$ case.

\section{Topology and $\theta$ dependence in $SU(N)$ gauge theories}
\label{sec:topo}

In this section we will summarize, for the convenience of the reader,
some basics facts about topology and $\theta$ dependence in $SU(N)$ theories, 
in such a way to make the comparison with the $G_2$ case simpler.

The euclidean Lagrangian density of the $SU(N)$ (continuum) gauge theory is
\begin{equation}\label{eq:L}
\mathcal{L}_{\theta}=\frac{1}{2}\mathrm{Tr}[F_{\mu\nu}F_{\mu\nu}]-i\theta q(x)\ , \qquad  
q(x)=\frac{g^2}{32\pi^2} \epsilon_{\mu\nu\rho\sigma}\mathrm{Tr}\Big[ F_{\mu\nu}(x) F_{\rho\sigma}(x)\Big] 
\end{equation}
and the associated free energy density can be computed by means of the relation
\begin{equation}\label{eq:F}
F(\theta, T)=-\frac{1}{V_4}\log\int[\mathscr{D}A]\exp\left(-\int_0^{1/T}\mathrm{d}t
\int_V\mathrm{d}^3 x\,\mathcal{L}_{\theta}\right)\ , \qquad V_4=T/V\ ,
\end{equation}
where $T$ is the temperature and $V$ the spatial volume. The topological charge
$Q=\int q(x)\mathrm{d}^4 x$ is odd under parity transformation, thus at
$\theta=0$ we have $\langle Q^{2n+1}\rangle_{\theta=0}=0$, since by the
Vafa-Witten theorem parity cannot be spontaneously broken \cite{Vafa:1984xg}.
As a consequence, assuming analyticity in $\theta=0$, $F(\theta, T)$ is an even
function of $\theta$ and can be expanded in the following form
\cite{Vicari:2008jw} 
\begin{equation}\label{eq:F_theta}
F(\theta,T)-F(0,T)=\frac{1}{2}\chi(T)\theta^2
\left[1+b_2(T)\theta^2+b_4(T)\theta^4+\cdots\right]\ .
\end{equation}
The topological susceptibility $\chi(T)$ and the coefficients $b_{2n}$ can be
computed by using the momenta of the topological charge distribution at $\theta=0$ as
\begin{equation}\label{eq:momenta_theta_zero}
\begin{aligned}
\chi(T)=\frac{\langle Q^2\rangle_{\theta=0}}{V_4}\qquad 
b_2=-\frac{\langle Q^4\rangle_{\theta=0}-
         3\langle Q^2\rangle^2_{\theta=0}}{12\langle Q^2\rangle_{\theta=0}}\\
b_4=\frac{\langle Q^6\rangle_{\theta=0}-15\langle Q^2\rangle_{\theta=0}\langle Q^4\rangle_{\theta=0} + 
    30\langle Q^2\rangle^3_{\theta=0}}{360\langle Q^2\rangle_{\theta=0}}\ .
\end{aligned}
\end{equation}

At finite temperature, instanton calculus does not suffer from infrared
divergences and can be used to gain some insight into the functional form of
$F(\theta, T)$.  The idea of the dilute instanton gas approximation is to replace the
path-integral expression of the partition function by the sum over an ensamble
of noninteracting instantons and anti-instantons. Denoting by $D^{-1/4}$ the
typical size of an instanton, we get \cite{Gross:1980br}
\begin{equation}\label{eq:inst_gas_Z}
\begin{aligned}
& \int[\mathscr{D}A]\exp\left(-\int_0^{1/T}\mathrm{d}t 
   \int_V\mathrm{d}^3 x\,\mathcal{L}_{\theta}\right)\approx \\
& \approx \sum_{n_+, n_-=0}^{\infty}\frac{1}{n_+!n_-!}(V_4D)^{n_+ + n_-} 
   \exp\left(-\frac{8\pi^2}{g^2}(n_++n_-)+i\theta(n_+-n_-)\right)= \\
&=\exp\left[2(V_4D)e^{-8\pi^2/g^2}\cos\theta\right]\ ,
\end{aligned}
\end{equation}
and thus (using the one loop running coupling constant and $D\sim T^4$)
\begin{align}
& F(\theta, T)-F(0, T)=\chi(T)(1-\cos\theta) \label{eq:inst_gas_F} \\
& \chi(T)\sim T^4\exp\big[-8\pi^2/g^2(T)]\sim T^{-\frac{11}{3}N+4} \label{eq:inst_gas_chi}\\
& b_2=-\frac{1}{12}\qquad b_4=\frac{1}{360}\qquad b_{2n}=(-1)^n\frac{1}{(2n+2)!}\ .  \label{eq:inst_gas_b}
\end{align}
The approximation in Eq.~\eqref{eq:inst_gas_Z} is expected to be reliable at
high temperatures; in this regime the instanton gas predicts a strong
suppression of the topological susceptibility, which gets stronger when
increasing the number of colors. This is in fact what is observed in numerical
simulations \cite{Alles:1996nm, Alles:1997qe, Alles:2000cg, Gattringer:2002mr,
Del Debbio:2004rw, Lucini:2004yh}: the topological susceptibility stays
constant for $T\lesssim T_c$, while it drops toward zero at deconfinement, in
qualitative accordance with Eq.~\eqref{eq:inst_gas_chi}. 

It has to be stressed that Eq.~\eqref{eq:inst_gas_chi} involves two different
approximations: the instanton gas approximation and the perturbative one, thus
it cannot be expect to be valid in the strongly coupled region near
deconfinement. Only the instanton gas approximation is instead used to obtain
the $b_{2n}$ values in Eq.~\eqref{eq:inst_gas_b}.

Another approach that give us some information on $F(\theta, T)$ is the 't
Hooft large $N$ limit \cite{'tHooft:1973jz, Lucini:2012gg}, which is expected
to be reliable at low temperature. If we do not want the $\theta$ dependence to
be washed out by the $N\to\infty$ limit, we have to impose that the two terms in
the Lagrangian Eq.~\eqref{eq:L} scale in the same way with $N$. Remembering that 
$g^2=\mathcal{O}(1/N)$, we obtain the large $N$ scaling form of the free energy 
\cite{Witten:1980sp, Witten:1998uka}:
\begin{equation}
F(\theta, T)=N^2\mathcal{F}(\bar{\theta}, T), \qquad \bar{\theta}=\theta/N\ .
\end{equation}
By comparing the power series in $\theta$ of the left and right hand side we get
\begin{equation}\label{eq:large_N_ris}
\chi(T)=\bar{\chi}(T)+\mathcal{O}(1/N^2)\qquad b_{2n}(T)=\bar{b}_{2n}(T)/N^{2n}\ ,
\end{equation}
where $\bar{\chi}$ and $\bar{b}_{2n}$ are the coefficient of the expansion of
$\mathcal{F}$ in power of $\bar{\theta}$ analogous to Eq.~\eqref{eq:F_theta}.
Several lattice measurements of $\chi(T)$ and $b_2$ exist at $T=0$ 
\cite{DelDebbio:2002xa, D'Elia:2003gr, Giusti:2007tu, Panagopoulos:2011rb}, 
that nicely follow the large $N$ scaling Eq.~\eqref{eq:large_N_ris}.

The functional dependence of the free energy on $\theta$ is expected to be
completely different in the low and high temperature phases: at low temperature
the natural variable is $\bar{\theta}=\theta/N$ and this, together with the
$2\pi$ periodicity in $\theta$, suggests $F$ to be a multibranched function of
the form \cite{Witten:1980sp, Witten:1998uka}
\begin{equation}\label{eq:F_branch}
F(\theta, T)=N^2\min_k H\left(\frac{\theta+2\pi k}{N}, T\right) \ .
\end{equation}
On the other hand, instanton gas predicts the form Eq.~\eqref{eq:inst_gas_F} of
the free energy, which is thus expected to be analytic in $\theta$ in the high
temperature phase. 

Although large $N$ methods and instanton calculus are reliable for $T\ll T_c$
and $T\gg T_c$ respectively, it appears natural to guess that the change of
regime happens exactly at deconfinement. In order to clarify this issue, in
\cite{Bonati:2013tt} the behaviour of $b_2$ across the deconfinement transition
was numerically investigated for $SU(3)$ and $SU(6)$ LGTs.
The advantage of $b_2$ with respect to the topological susceptibility
$\chi(T)$ is that we have a clear cut distinction between the two possible
behaviours: $b_2$ scales as $1/N^2$ or is independent of $N$.  Moreover, as
previously noticed, the value of $b_2$ in the high temperature regime is unambiguously
predicted by the instanton gas approximation, all the uncertainties related to
perturbation theory being factorized into $\chi(T)$.
The value of $b_2$ observed at $T\approx 0.95 T_c$ is compatible with the one
at $T=0$ and scales according to Eq.~\eqref{eq:large_N_ris}; above the
transition ($T\gtrsim 1.05T_c$) the value of $b_2$ does not scale with $N$,
thus indicating that the relevant variable is not $\bar{\theta}$ but just $\theta$.
The instanton gas prediction for $b_{2}$ and $b_4$ turned out to be well
satisfied for temperature just slightly above deconfinement ($T\gtrsim 1.1T_c$).
These properties are also reproduced by model calculations in QCD-like theories
\cite{Poppitz:2012nz, Bergman:2006xn, Parnachev:2008fy, Thomas:2011ee}.

\section{$G_2$ lattice gauge theory}
\label{sec:G_2}

In LGT \cite{Wilson:1974sk} the elementary objects are the parallel transports
along the links of the lattice, that will be denoted by $U_{\mu}(x)$ and are
elements of the gauge group. 
The group $G_2$ can be identified with the group of the automorphism of the
octonions \cite{AdamsBook}, which is isomorphic to the subgroup of $SO(7)$ that
leaves invariant a specific $3-$form \cite{ReeseHarveyBook}: a $7\times 7$ real
matrix $M$ is an element of $G_2$ if and only if $M\in SO(7)$ and
\begin{equation}
T_{abc}=T_{a'b'c'}M_{aa'}M_{bb'}M_{cc'}\ ,
\end{equation}
where $T_{abc}$ is the completely antisymmetric tensor whose non-vanishing
elements (up to permutations) are given by\footnote{The explicit form of the
$T$ tensor is base dependent, for other possible choices see, e.g.
\cite{Greensite:2006sm}.} \cite{Cossu:2007dk} 
\begin{equation}
T_{123} = T_{176} = T_{145} = T_{257} = T_{246} = T_{347} = T_{365} = 1.
\end{equation}
In our simulations we adopted the standard Wilson plaquette action
\cite{Wilson:1974sk}
\begin{equation}\label{eq:w_action}
S_W=-\beta\sum_{x}\,\sum_{0\le \mu,\nu\le 3} \mathrm{Tr} P_{\mu\nu}(x)\ ,
\end{equation}
where $P_{\mu\nu}$ is the product of four links around an elementary plaquette:
\begin{equation}
P_{\mu\nu}(x)=U_{\mu}(x)U_{\nu}(x+a\hat{\mu})U_{\mu}^{\dag}(x+a\hat{\nu})U_{\nu}^{\dag}(x)\ .
\end{equation}
Notice that in the literature two different conventions on the value of $\beta$
exist for the $G_2$ theory: the one in Eq.~\eqref{eq:w_action} and the one that
corresponds to the change $\beta\to\beta/7$ in Eq.~\eqref{eq:w_action}.

A particularly convenient basis for the $G_2$ algebra was constructed in
\cite{Cacciatori:2005yb}, which can be adopted to easily identify the $SU(2)$
subgroups to be used in a Monte-Carlo update \emph{a la} Cabibbo-Marinari
\cite{Cabibbo:1982zn}. The application of this method to the $G_2$ theory is
not, however, completely trivial: as can be explicitly seen by looking at the
expressions in App. A of \cite{Greensite:2006sm}, only three $SU(2)$ subgroups
are embedded in $G_2$ in a way simple enough to be efficiently used in a
standard heatbath/overrelaxation update \cite{Creutz:1980zw, Kennedy:1985nu,
Creutz:1987xi}. These $SU(2)$ subgroups do not cover completely the $G_2$ group
and, to ensure the ergodicity of the update algorithm, random gauge
transformations has to be also applied.

Before going on, a small digression is necessary on the normalization of the
topological charge. Instantons and topological charge are usually discussed in
the setting of $SU(N)$ gauge theories, in which the topological charge density
is given by Eq.~\eqref{eq:L} and the normalization is fixed by the requirement
that the topological charge $Q=\int q(x)\mathrm{d}x$ has to be equal to the 
winding number associated to $\pi_3(SU(N))=\mathbb{Z}$ (see e.g.
\cite{EWeinbergBook, SWeinbergBook}). Use of the expression Eq.~\eqref{eq:L}
in the $G_2$ gauge theory would however give only even topological charges.  The
correct normalization to be used in the general case has been discussed in
\cite{Bernard:1977nr} and the final result is
\begin{equation}\label{eq:q_general}
q(x)=\frac{g^2}{64 K \pi^2} \epsilon_{\mu\nu\rho\sigma}
\mathrm{Tr}\Big[ F_{\mu\nu}(x) F_{\rho\sigma}(x)\Big]\ ,
\end{equation}
where the algebra generators $T^a$s are normalized in such a way that the
longest root is equal to $1$ and $K$ is given by the relation
$\mathrm{Tr}(T^aT^b)=K\delta^{ab}$. Using the explicit realization given in
\cite{Cacciatori:2005yb} of the $G_2$ algebra it is simple to show that for
$G_2$ we have $K=1$ (while $K=1/2$ for $SU(N)$). 

On the lattice several methods exist to associate a value of the topological
charge to a given configuration. Since all these methods has been proven to
give equivalent results (see e.g \cite{Vicari:2008jw, Bonati:2014tqa}) and high
statistics is needed in our study, we adopted the cheapest from the numerical
point of view. The topological charge has been measured after cooling and the
simplest discretization with definite parity of Eq.~\eqref{eq:q_general} was
adopted, namely \cite{Di Vecchia:1981qi} 
\begin{equation}\label{qL}
q_L(x) = -\frac{1}{2^4\times 64 K \pi^2}\sum_{\mu\nu\rho\sigma = \pm 1}^{\pm 4} 
\tilde{\epsilon}_{\mu\nu\rho\sigma} \mathrm{Tr} \left(U_{\mu\nu}(x) U_{\rho\sigma}(x) \right) \; ,
\end{equation}
where ${\tilde{\epsilon}}_{\mu\nu\rho\sigma}$ coincides with the usual
Levi-Civita tensor for positive indices, while for the negative directions it
is defined by the relation ${\tilde{\epsilon}}_{\mu\nu\rho\sigma} =
-{\tilde{\epsilon}}_{(-\mu)\nu\rho\sigma}$ and the complete antisymmetry.  This
approach is the same adopted e.g. in \cite{Bonati:2014tqa}, to which we refer
for technical details, the only difference being that, for the $G_2$ gauge
theory, cooling consists of cooling on three $SU(2)\subset G_2$ and random
gauge transformations.

\section{Numerical results}
\label{sec:num_res}

In this section we will present the numerical results obtained by means of
Monte-Carlo simulations of the $G_2$ lattice theory.  Most of the finite
temperature simulations have been performed by using lattices with temporal
extent\footnote{This is the smallest $N_t$ value for which the finite
temperature transition takes place at a critical coupling larger than the one
corresponding to the bulk transition, see e.g.  \cite{Cossu:2007dk}.} $N_t=6$;
by using three different values for the spatial size, $N_s=12, 18, 24$, we
verified that an aspect ratio $N_s/N_t=3$ is sufficient to neglect finite size
effects with our statistical uncertainties.  To check for the continuum limit a
$8\times 24^3$ lattice was then simulated.  

\begin{figure}[t]
\centering 
\includegraphics*[width=.7\textwidth]{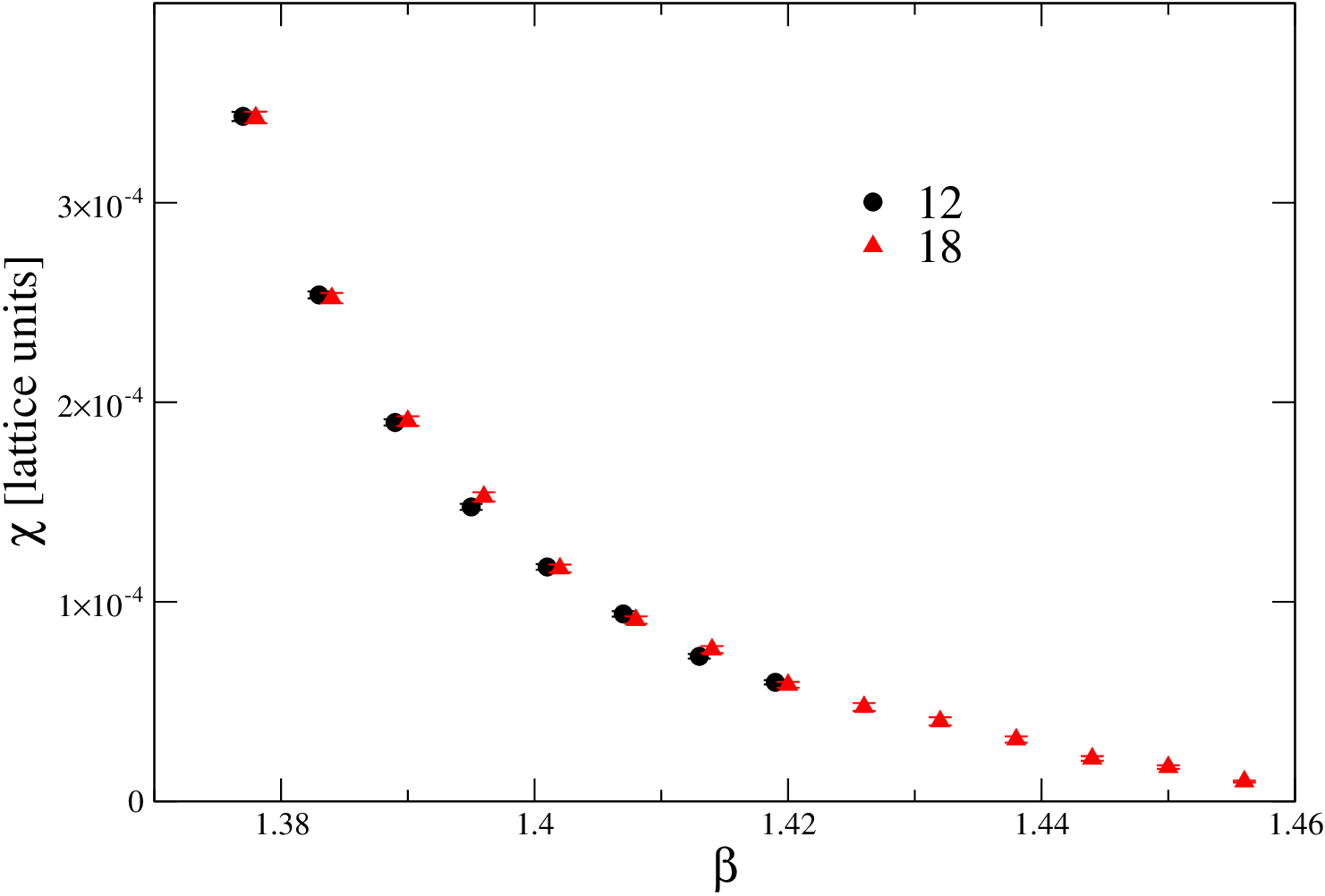}
\caption{Zero temperature topological susceptibility measured on
$12^4$ an $18^4$ lattices. Data measured on the $18^4$ lattice have been
slightly shifted to the right for clarity.} \label{fig:chi_0}
\end{figure}

The results reported in the following have been obtained by using the
topological charge extracted after $90$ cooling steps (see
\cite{Bonati:2014tqa} for more details on the procedure adopted), but they
proved to be stable within errors for a number of cooling steps in the range
from $30$ to $150$. A statistics of $\mathcal{O}(10^5)$ measures have been used for
each coupling value, with measures performed every $10$ update steps and each
step consisting of a Cabibbo-Marinari heatbath, five Cabibbo-Marinari
overrelaxations and a random gauge transformation.

The values of $T/T_c$ have been estimated by using the parametrization reported
in \cite{Bruno:2014rxa} for the string tension and the following values of
the critical couplings: $\beta_c(N_t=6)=1.3951(2)$ (see \cite{Cossu:2007dk}) and
$\beta_c(N_t=8)=1.431(3)$ (compatible with the one reported in
\cite{Bruno:2014rxa}).

\begin{figure}[t]
\centering 
\includegraphics*[width=.45\textwidth]{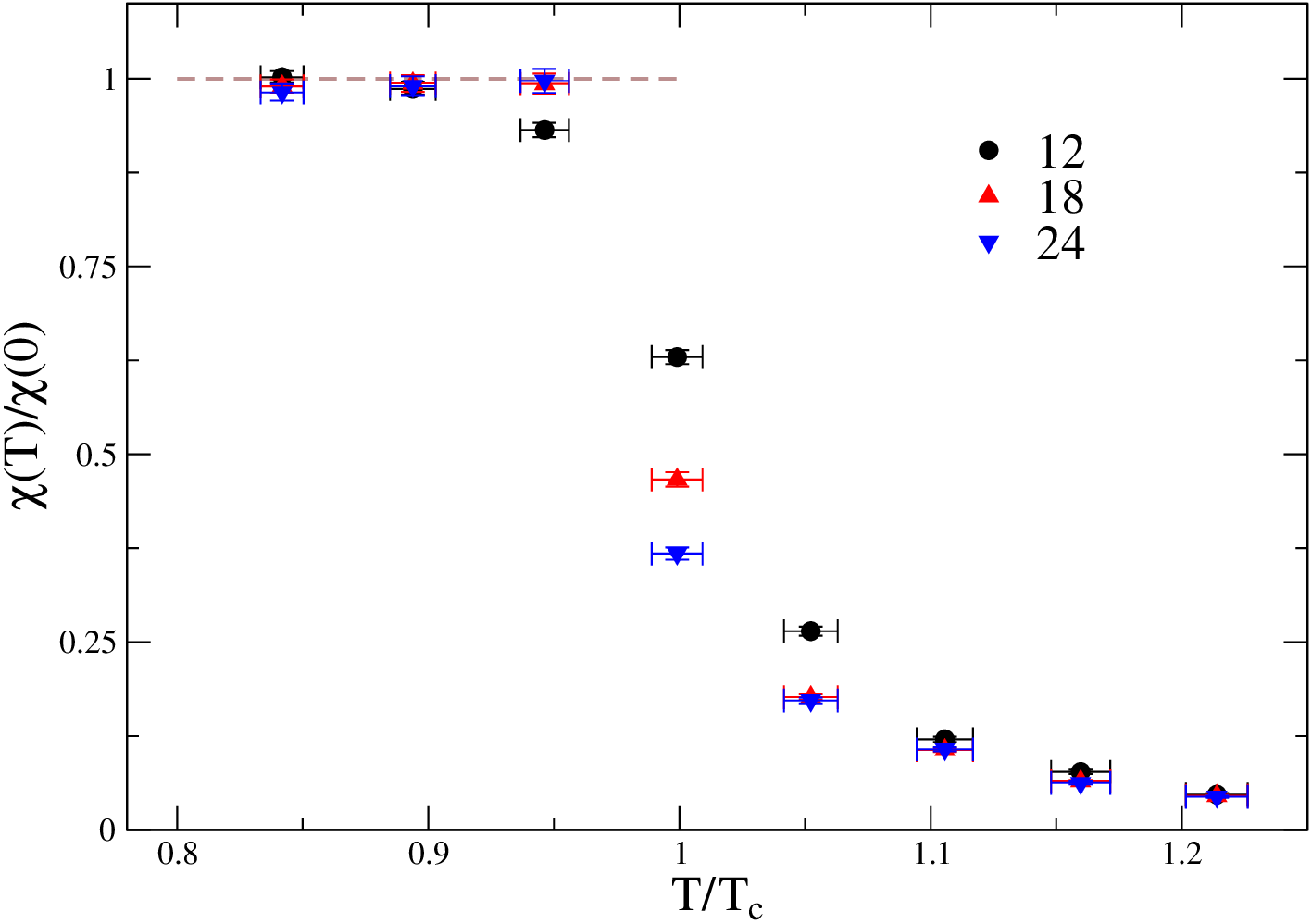}
\hspace{0.5cm}
\includegraphics*[width=.45\textwidth]{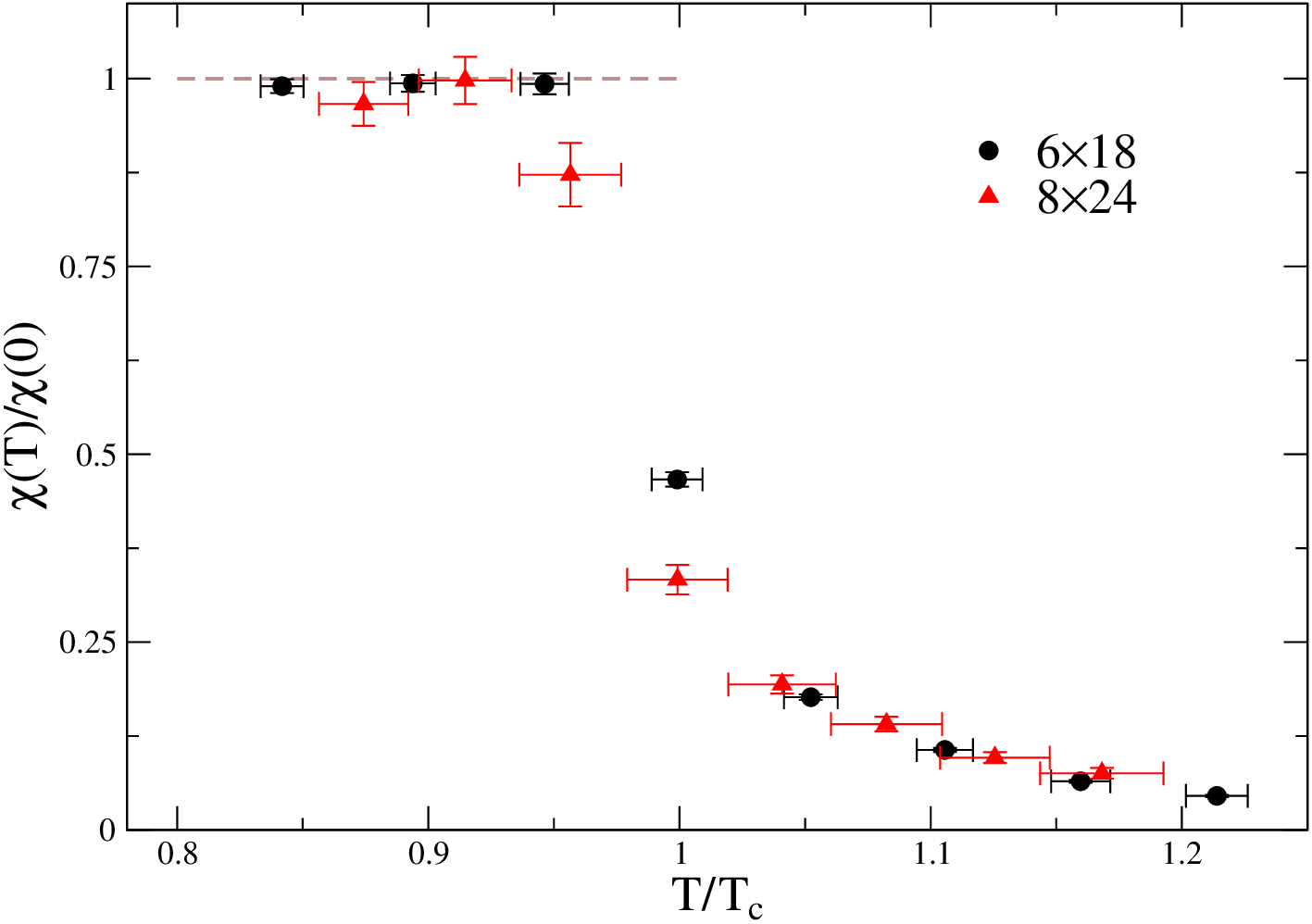}
\caption{Ratio of the finite temperature topological susceptibility and the
zero temperature one.  (\emph{Left}) Results for the lattices $6\times N_s^3$,
with $N_s=12, 18, 24$.  (\emph{Right}) Comparison of the results obtained on
$6\times 18^3$ and $8\times 24^3$ lattices.}\label{fig:chi_ratio}
\end{figure}

The first observable studied has been the topological susceptibility $\chi(T)$
and, in particular, the dimensionless ratio $\chi(T)/\chi(T=0)$. In order to
compute this ratio simulations have been performed on symmetric lattices at the
same coupling values adopted for the finite temperature runs. The
results obtained for $\chi(T=0)$ on the lattices $12^4$ and $18^4$ are shown in
Fig.~\eqref{fig:chi_0}. From the comparison of the results obtained on the two
lattices we can conclude that the $12^4$ data do not show significant finite
size effects. Moreover the size of the $18^4$ lattice at $\beta=1.455$
($\approx 1.5\,\mathrm{fm}$) is larger than the one of the $12^4$ lattice at
$\beta=1.419$ ($\approx 1.3\,\mathrm{fm}$); we thus conclude that also the
$18^4$ data are not affected by significant finite size effects.

In Fig.~\eqref{fig:chi_ratio} the estimated values of the ratio
$\chi(T)/\chi(0)$ are shown. The left panel displays the results obtained on
the $N_t=6$ lattices for several spatial extents and, as previously
anticipated, no significant finite size effects can be seen as far as the
aspect ratio is $3$ or larger (apart from the data at $T\approx T_c$). In the
right panel of Fig.~\eqref{fig:chi_ratio} we compare the results obtained by
using two different lattice spacing at fixed physical lattice size. The nice
agreement between the determinations obtained by using the $6\times 18^3$ and
the $8\times 24^3$ lattices supports the absence of significant lattice
artefacts in our measurements. We can thus conclude that, like in $SU(N)$ gauge
theories, $\chi(T)$ stays constant for $T<T_c$, with an abrupt decrease at
deconfinement.

\begin{figure}[t]
\centering 
\includegraphics*[width=.7\textwidth]{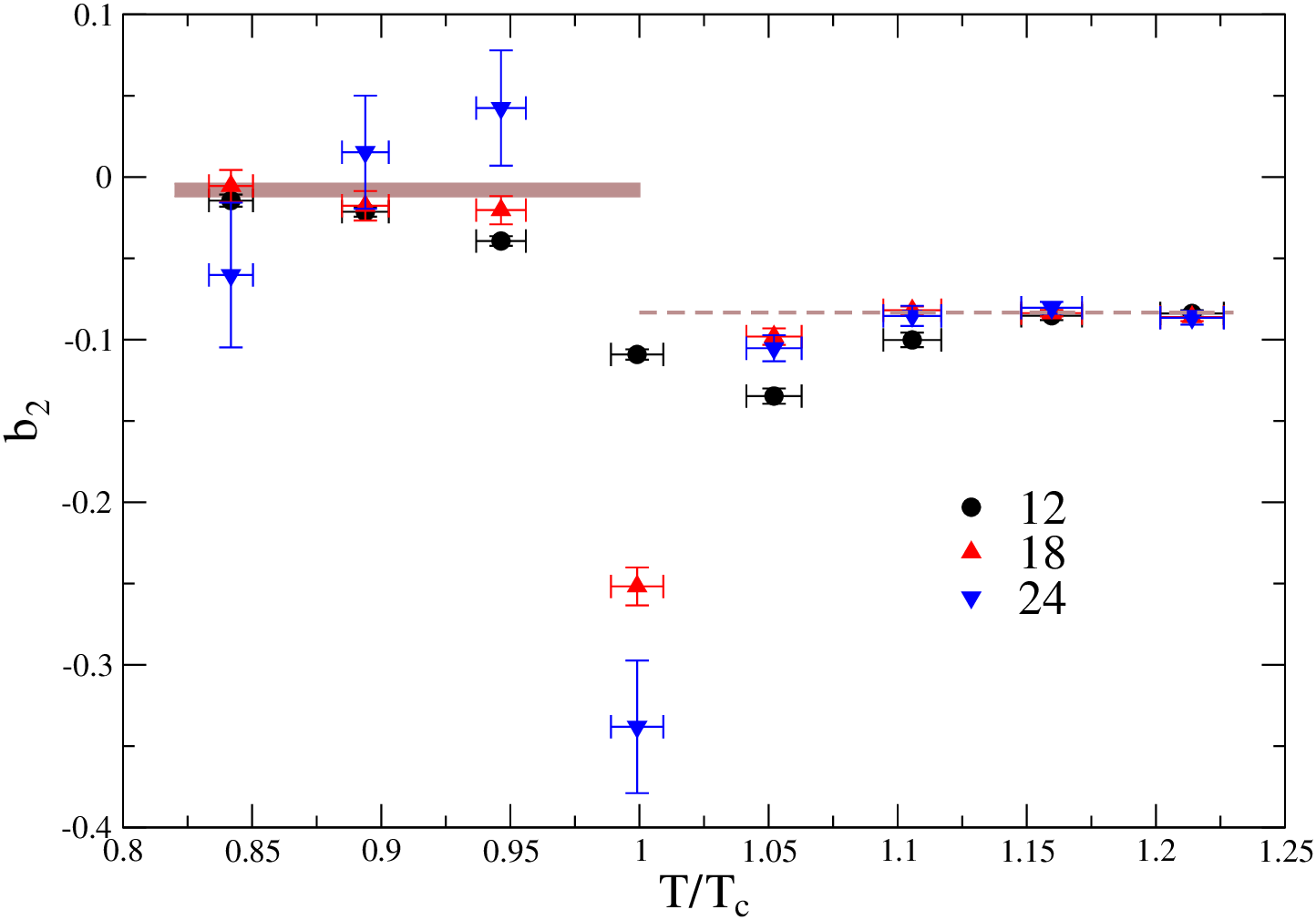}
\caption{Plot of the $b_2$ value across the deconfinement temperature for
lattices $6\times N_s^3$, with $N_s=12, 18, 24$. The band for $T/T_c\le 1$ 
is the $b_2$ value for $SU(6)$ (see \cite{Bonati:2013tt}), while the 
dashed line for $T/T_c>1$ denotes the dilute instanton gas value 
$b_2^{\mathrm{(dig)}}=-1/12$.}
\label{fig:b2_large}
\end{figure}

\begin{figure}[b]
\centering 
\includegraphics*[width=.45\textwidth]{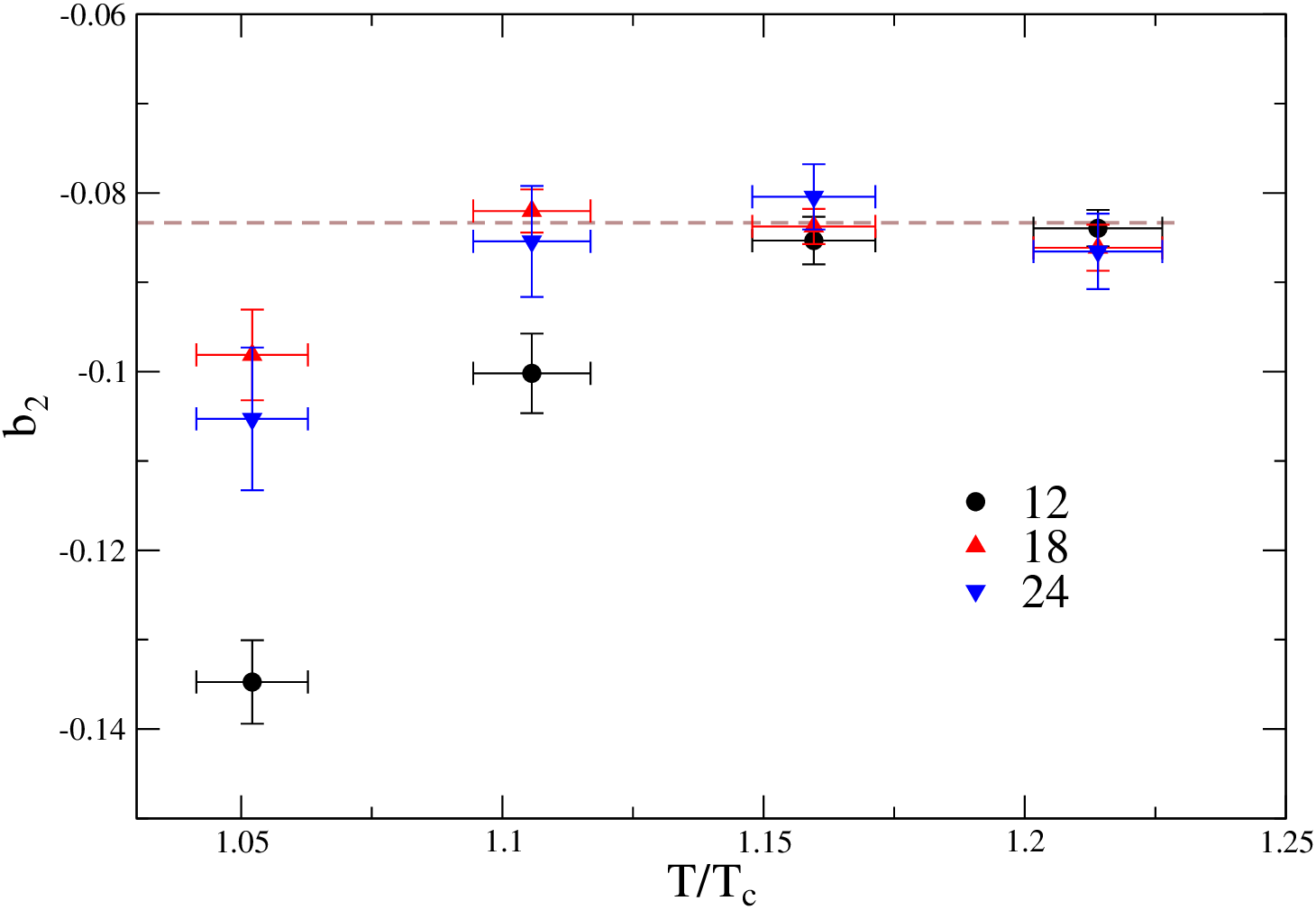}
\hspace{0.5cm}
\includegraphics*[width=.45\textwidth]{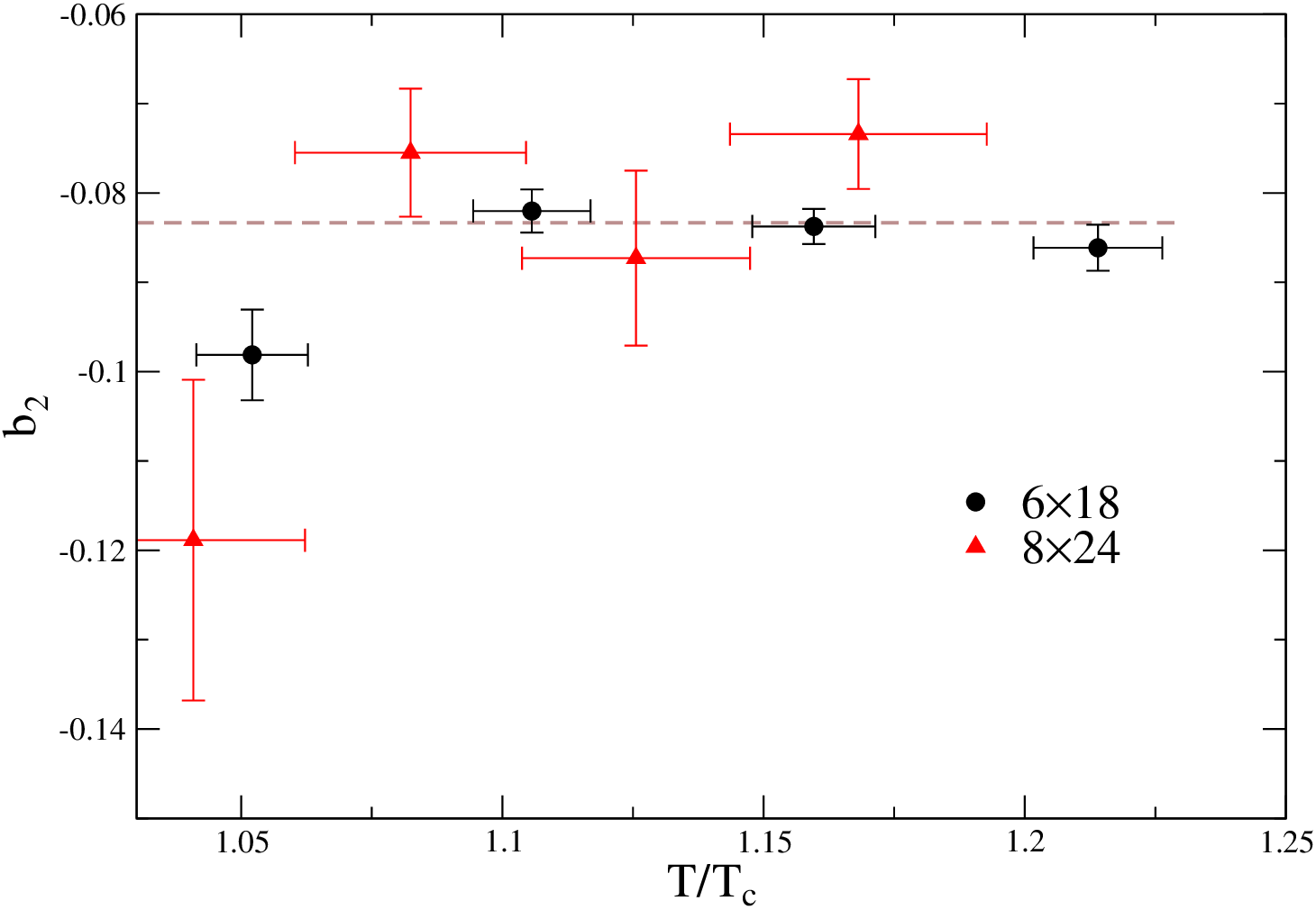}
\caption{$b_2$ value in the high temperature region. The horizontal dashed line
denotes the dilute instanton gas result $b_2^{\mathrm{(dig)}}=-1/12$.
(\emph{Left}) Results for the lattices $6\times N_s^3$, with $N_s=12, 18, 24$.
(\emph{Right}) Comparison of the results obtained on $6\times 18^3$ and
$8\times 24^3$ lattices.}
\label{fig:b2_small}
\end{figure}

The value of the parameter $b_2$ across the deconfinement transition is shown
in Fig.~\eqref{fig:b2_large} for the $N_t=6$ lattices. As a first observation
we notice that, although the statistics (and the autocorrelations) are of the
same order of magnitude for all the lattice sizes studied, the error bars of
the low temperature data for the $6\times 24^3$ lattice are much larger that
the ones for the smaller lattices. This phenomenon is known to happen also in
$SU(N)$ LGTs and is likely related to the peculiar form
Eq.~\eqref{eq:momenta_theta_zero} of the $b_{2n}$ observables when computed by
means of simulations at $\theta=0$.  A possible way out could be to perform
simulations at imaginary value of the $\theta$ parameter, as proposed in
\cite{Panagopoulos:2011rb}, however it would be difficult to study in this way
the neighbourhood of the deconfinement transition, since the value of the
deconfinement temperature $T_c$ also depend on $\theta$ (see
\cite{D'Elia:2012vv, D'Elia:2013eua} for lattice studies and
\cite{Unsal:2012zj,Anber:2013sga, Bigazzi:2014qsa} for other approaches).
Since for all the more standard observables the results obtained on the lattice
$6\times 18^3$ are completely equivalent (apart from the $T\approx T_c$ data)
to the ones obtained on the larger $6\times 24^3$ lattice, we expect that, also
for $b_2$, data obtained on the lattice with aspect ratio $3$ do not suffer
from severe finite size effects and we will refer to them in the following
discussions.

Let us now come to the behaviour of $b_2$ across the transition.  In the low
temperature phase the value of $b_2$ is quite small, almost compatible with
zero, much like what happens in $SU(N)$ gauge theories; for reference the value
of $b_2$ for $SU(6)$ at zero temperature ($b_2=0.008(4)$, \cite{Bonati:2013tt})
is also shown in Fig.~\eqref{fig:b2_large}. For $T>T_c$ the value of $b_2$
increases by an order of magnitude and promptly approaches the asymptotic value
predicted by the dilute instanton gas, $b_2^{\mathrm{(dig)}}=-1/12$. To better
appreciate the rapidity of the convergence to the asymptotic value, in
Fig.~\eqref{fig:b2_small} we display a magnification of the high temperature
region, both for $N_t=6$ and $N_t=8$ lattices. Significant deviations from the
value $b_2^{\mathrm{(dig)}}$ are visible only for $T\lesssim 1.1T_c$.

\begin{figure}[t]
\centering 
\includegraphics*[width=.45\textwidth]{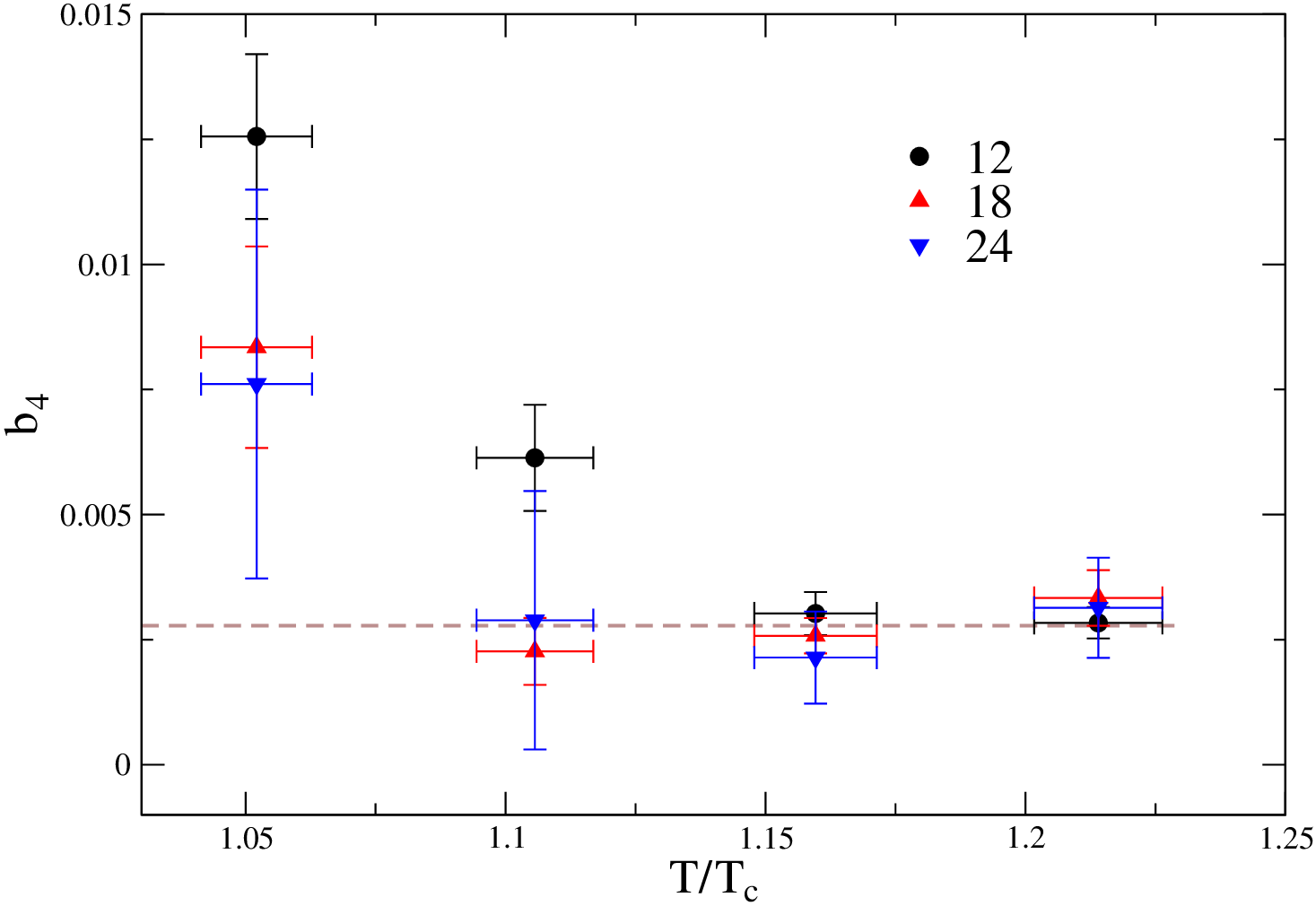}
\hspace{0.5cm}
\includegraphics*[width=.45\textwidth]{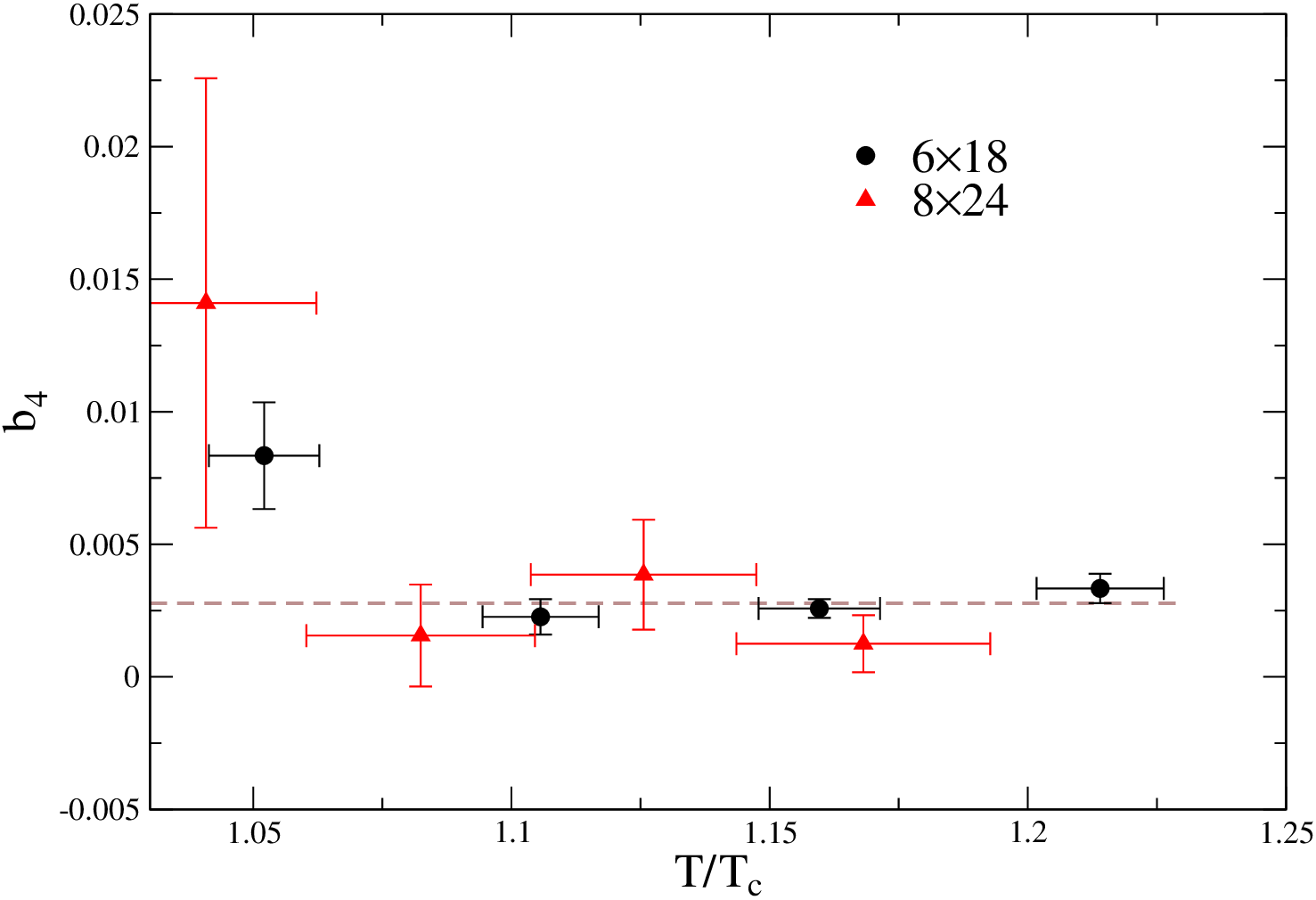}
\caption{$b_4$ value in the high temperature region. The horizontal dashed line
denotes the dilute instanton gas result $b_4^{(\mathrm{dig})}=1/360$.
(\emph{Left}) Results for the lattices $6\times N_s^3$, with $N_s=12, 18, 24$.
(\emph{Right}) Comparison of the results obtained on $6\times 18^3$ and
$8\times 24^3$ lattices.}
\label{fig:b4_small}
\end{figure}

The numerical estimates of $b_4$ in the low temperature phase are too noisy to
extract from them useful informations, but high temperature data are precise
enough to show the convergence to the dilute instanton gas value
$b_4^{\mathrm{(dig)}}=1/360$, see Fig.~\eqref{fig:b4_small}. Like for the case
of $b_2$, no significant discrepancies from the asymptotic value
$b_4^{\mathrm{(dig)}}$ can be see as far as $T\gtrsim 1.1T_c$.

\section{Conclusions}
\label{sec:concl}

In this work we studied the behaviour, across the deconfinement transition, of 
the topological susceptibility $\chi(T)$ and of the coefficients $b_{2n}(T)$ that
parametrize the $\theta$ dependence of the free energy in the $G_2$ gauge
theory. 

The sudden drop at deconfinement of the ratio $\chi(T)/\chi(T=0)$ signals that
in the high temperature phase the topological activity is strongly suppressed.
The abrupt change of the $b_2$ coefficient at $T_c$ shows that the difference
between the low and the high temperature phases is not just a difference in the
global activity, but that also the functional form of the $\theta$ dependence
of the free energy has changed. The $b_2$ and $b_4$ values in the deconfinded
phase rapidly approach the predictions of the dilute instanton gas model, which
well reproduce lattice data for $T\gtrsim 1.1 T_c$.

The picture that emerges is surprisingly similar to that of the $SU(N)$ gauge
theory, with however a fundamental difference: in the $SU(N)$ setup the change
of the $\theta$ dependence at deconfiement can be conveniently interpreted as a
change from a low temperature large-$N$ regime to an high temperature instanton
one.  While the argument for the instanton-like behaviour of the free energy in
the deconfined phase can be applied without modifications to the $G_2$ gauge
theory, it is not clear what takes the role of the large-$N$ regime for this
theory.  

The most natural explanation of this common behaviour of $SU(N)$ and $G_2$
theories is probably that the confinement mechanism is the same for all the
simple gauge groups and that the degrees of freedom responsible for the
confinement have non-trivial topological properties. Several proposals that go
in this directions exist in the literature and are actively investigated, like
dyons \cite{Diakonov:2010qg}, bions \cite{Poppitz:2012nz, Anber:2014lba,
Unsal:2012zj}, instanton-quarks \cite{Parnachev:2008fy, Zhitnitsky:2008ha} and
the relations between monopoles and instantons \cite{DiGiacomo:2015eva}, not to
mention the analogy between the $\theta$ angle and the chemical potential noted
in \cite{D'Elia:2013eua} and the similarities with spin models
\cite{Bruckmann:2007zh, Dunne:2012ae, Cherman:2013yfa, Bruckmann:2014sla}.

A related point is the global analytical structure of the free energy as a
function of $\theta$ in the confined phase. We previously recalled the
large-$N$ Witten's argument, which suggests that the free energy of the $SU(N)$
theory is a multi-branched function of the form Eq.~\eqref{eq:F_branch} and
gives a qualitative explanation of the smallness of $b_2$ for $T<T_c$.  
It is tempting to relate the small value of $b_2$ observed also in the
low temperature phase of the $G_2$ theory to an analogous multi-branched
structure of the free energy, which however have no natural large-$N$
interpretation.
 
The early onset of the dilute instanton gas regime, just slightly above the
deconfinement transition, also appears to be a feature common to both $SU(N)$
and $G_2$ gauge theories, with indications that this is true also in presence
of quarks \cite{Cossu:2013uua, Kanazawa:2014cua}. While it is not clear if
these two common features, i.e. the change of $\theta$ dependence at
deconfinement and the early onset of the instanton behaviour, are related to
each other or not, it is interesting to notice that also deviations from the
dilute instanton gas behaviour are qualitatively similar in $SU(N)$ and $G_2$
gauge theories, with $b_2$ approaching its asymptotic value from below and
$b_4$ from above.

\acknowledgments

It is a pleasure to thank Philippe de Forcrand, Massimo D'Elia and Marco Panero
for useful discussions and comments.  Numerical simulations have been performed
using GRID resources provided by the Scientific Computing Center at INFN-Pisa.

\appendix

\section{Numerical data}

In Tabs.~\eqref{tab:6x12},\eqref{tab:6x18},\eqref{tab:6x24} and
\eqref{tab:8x24} the values of the bare coupling used for the various lattices
and the estimates obtained for $T/T_c$, $\chi(T)/\chi(T=0)$, $b_2$ and $b_4$
are reported.

\begin{table}
\centering
\begin{tabular}{lllll}
$\beta$ & $T/T_c$   & $\chi(T)/\chi(0)$ & $b_2$ & $b_4$ \\ \hline
1.377 &  0.8418(85)  &  1.0021(80) &  -0.0145(37)  &   -0.0022(23) \\ \hline
1.383 &  0.8938(91)  &  0.9866(83) &  -0.0214(31)  &   -0.0007(16) \\ \hline
1.389 &  0.9463(96)  &  0.9317(95) &  -0.0393(30)  &   -0.0006(18) \\ \hline
1.395 &  0.999(10)   &  0.6294(93) &  -0.1092(32)  &    0.0008(11) \\ \hline
1.401 &  1.052(11)   &  0.2644(60) &  -0.1347(47)  &    0.0126(16) \\ \hline
1.407 &  1.106(11)   &  0.1206(35) &  -0.1002(44)  &    0.0061(11) \\ \hline
1.413 &  1.160(12)   &  0.0773(28) &  -0.0853(27)  &    0.00302(43) \\ \hline
1.419 &  1.214(12)   &  0.0469(20) &  -0.0839(20)  &    0.00284(31)  \\ \hline
\end{tabular}
\caption{Data for the lattice $6\times 12^3$.}\label{tab:6x12}
\end{table}

\begin{table}
\centering
\begin{tabular}{lllll}
$\beta$ & $T/T_c$   & $\chi(T)/\chi(0)$ & $b_2$ & $b_4$ \\ \hline
1.377 &  0.8418(85) &  0.9901(91)  &  -0.0055(98)  &   -0.034(18)  \\ \hline
1.383 &  0.8938(91) &  0.994(11)   &  -0.0177(90)  &    0.019(15)   \\ \hline
1.389 &  0.9463(96) &  0.993(14)   &  -0.0204(86)  &    0.0059(97)  \\ \hline
1.395 &  0.999(10)  &  0.4664(97)  &  -0.252(12)   &    0.045(14)   \\ \hline
1.401 &  1.052(11)  &  0.1767(36)  &  -0.0981(51)  &    0.0083(20)  \\ \hline
1.407 &  1.106(11)  &  0.1065(25)  &  -0.0820(24)  &    0.00227(67)  \\ \hline
1.413 &  1.160(12)  &  0.0647(18)  &  -0.0837(20)  &    0.00258(35)  \\ \hline
1.419 &  1.214(12)  &  0.0452(15)  &  -0.0861(26)  &    0.00333(56)  \\ \hline
\end{tabular}
\caption{Data for the lattice $6\times 18^3$.}\label{tab:6x18}
\end{table}

\begin{table}
\centering
\begin{tabular}{lllll}
$\beta$ & $T/T_c$   & $\chi(T)/\chi(0)$ & $b_2$ & $b_4$ \\ \hline
1.377 &  0.8418(85) &  0.982(11)  &  -0.060(45)   & -0.29(20)  \\ \hline
1.383 &  0.8938(91) &  0.990(13)  &   0.015(35)   & -0.15(11)  \\ \hline
1.389 &  0.9463(96) &  0.997(16)  &   0.042(35)   &  0.132(87) \\ \hline
1.395 &  0.999(10)  &  0.368(81)  &  -0.338(41)   &  0.270(88) \\ \hline 
1.401 &  1.052(11)  &  0.1721(38) &  -0.1053(80)  &  0.0076(39) \\ \hline
1.407 &  1.106(11)  &  0.1074(28) &  -0.0854(62)  &  0.0029(26) \\ \hline
1.413 &  1.160(12)  &  0.0623(19) &  -0.0804(37)  &  0.00214(92) \\ \hline
1.419 &  1.214(12)  &  0.0442(16) &  -0.0865(42)  &  0.0031(10) \\ \hline
\end{tabular}
\caption{Data for the lattice $6\times 24^3$.}\label{tab:6x24}
\end{table}

\begin{table}
\centering
\begin{tabular}{lllll}
$\beta$ & $T/T_c$   & $\chi(T)/\chi(0)$ & $b_2$ & $b_4$ \\ \hline
1.413 &  0.873(18)  &  0.966(30)   & -0.114(45)  &  0.065(72) \\ \hline
1.419 &  0.916(18)  &  0.997(31)   &  0.022(23)  & -0.025(20)  \\ \hline
1.425 &  0.956(20)  &  0.872(42)   & -0.053(22)  & -0.034(15) \\ \hline
1.431 &  0.997(20)  &  0.333(20)   & -0.097(13)  & -0.0041(35) \\ \hline
1.437 &  1.040(21)  &  0.194(12)   & -0.119(18)  &  0.0141(85) \\ \hline
1.443 &  1.082(22)  &  0.1409(96)  & -0.0755(71) &  0.0016(19) \\ \hline
1.449 &  1.125(22)  &  0.0962(73)  & -0.0873(98) &  0.0039(21) \\ \hline
1.455 &  1.168(25)  &  0.0753(72)  & -0.0734(41) &  0.0012(11)  \\ \hline
\end{tabular}
\caption{Data for the lattice $8\times 24^3$.}\label{tab:8x24}
\end{table}

\end{document}